     Based on talks presented by N. Mavromatos and D. Nanopoulos
at the International Symposium on Blackholes, Membranes, Wormholes,
and Superstrings, January 16-18 1992, HARC (Houston), USA.
\\
\documentstyle[12pt]{article}
\begin{document}
\newcommand{\vsone}{\vspace{1cm}}
\newcommand{\be}{\begin{equation}}
\newcommand{\ee}{\end{equation}}
\newcommand{\bea}{\begin{eqnarray}}
\newcommand{\eea}{\end{eqnarray}}
\newcommand{\pr}{\paragraph{}}

\newcommand{\nd}[1]{/\hspace{-0.6em} #1}
\begin{titlepage}
\begin{flushright}
CERN-TH.6476/92\\
ACT-7/92 \\
CTP-TAMU-28/92 \\
\end{flushright}

\begin{centering}
\vspace{.1in}
{\large {\bf On the $W$-hair  of String
Black Holes and the Singularity Problem }} \\

\vspace{.2in}
{\bf John Ellis} and {\bf N.E. Mavromatos}\\
\vspace{.05in}
Theory Division, CERN, CH-1211, Geneva 23, Switzerland  \\
and \\

\vspace{.05in}
{\bf D.V. Nanopoulos}\\

\vspace{.05in}
Center for Theoretical Physics, Dept. of Physics, \\
Texas A \& M University, College Station, TX 77843-4242, USA \\
\vspace{.03in}
and \\
\vspace{.05in}
Astroparticle Physics Group \\
Houston Advanced Research Center (HARC),\\
The Woodlands, TX 77381, USA\\
\vspace{.1in}
{\bf Abstract} \\
\vspace{.03in}
\end{centering}
{\small
\paragraph{}
We
argue
that the infinitely many
gauge symmetries of string theory provide an infinite set
of conserved (gauge)
quantum numbers ($W$-hair) which characterise
black hole states and maintain quantum coherence,
even during exotic processes like black hole
evaporation/decay.
We
study ways of measuring
the
$W$-hair of spherically-symmetric four-dimensional objects
with event horizons, treated as effectively
two-dimensional
string black holes.
Measurements can be done either through the
s-wave
scattering of
light particles
off the string black-hole background,
or
through interference
experiments of
Aharonov-Bohm
type.
We also
speculate on the r\^ ole of the extended
$W$-symmetries possessed
by the
topological
field theories that describe the
region of space-time around
a singularity.}
\par
\vspace{0.2in}
\begin{centering}

{\it Based on talks presented by N. Mavromatos and D. Nanopoulos
at the International Symposium on Blackholes, Membranes, Wormholes,
and Superstrings,
January 16-18 1992, HARC (Houston), USA.}

\end{centering}
\vspace{0.1in}
\begin{flushleft}
CERN-TH.6476/92 \\
ACT-7/92 \\
CTP-TAMU-28/92 \\
April 1992 \\
\end{flushleft}
\end{titlepage}
\newpage
\section{Introductory Review }
\paragraph{}
The reconciliation of general relativity with
quantum mechanics is one of the key problems
in physics. One of its many facets
is the quantisation of gravitational
effects in a flat space-time background, including
the calculability of perturbation theory, which should be
{\it finite} or at least {\it renormalisable}. Another
facet is the problem of quantisation in non-flat
space-times that are solutions of the gravitational
field equations. Still another is the full non-perturbative
treatment of fluctuations in space-time. And so on...
\pr
Conventional point-like quantum field theory has been
unable to resolve any of these problems, whilst string
theory is an ambitious candidate for a framework in which
they can all be resolved. Indeed, it has now been established
that string perturbation theory in a flat space-time background
is {\it finite} \cite{mand}. On the other hand, the full
non-perturbative treatment of fluctuations in space-time
requires the development of a fully-fledged string field
theory, which so far only exists for simplified toy models
of string gravity coupled to matter \cite{das,sen}.
In a series of recent papers \cite{emn1,emn2,emn3,emn4,emnl},
we have been studying the intermediate problem of quantum
theory in a non-flat space-time background.
\pr
Specifically, we have been investigating whether
conventional quantum coherence can be maintained in a black hole
background in string theory. Semiclassical arguments \cite{hawk}
in conventional
point-like quantum field theory indicate that macroscopic
black holes behave like mixed, thermal states, which has
motivated suggestions \cite{hawpop} that quantum coherence
{\it cannot} be maintained at a fundamental level
when microscopic non-perturbative
fluctuations in space-time are taken into account. The root
of this problem with quantum coherence is the observation
that the apparent entropy $S$ of a black hole is proportional
to the area of its event horizon \cite{hawk,bek}: $S \propto  A$,
which is in turn $\propto  M^2$ for an uncharged axisymmetric
black hole. Thus the apparent entropy of a black hole is
unbounded, whereas any point-like quantum field theory
has only a finite set
of quantum numbers. These are insufficient
to encode all the information carried by matter
that collapses gravitationally,
and thereby distinguish
all the states of a black hole, which must therefore be described
by a {\it mixed} state.
\pr
However, we have argued \cite{emn1,emn2}
that there are infinitely many {\it gauge} symmetries
in string theory, which lead to an {\it infinity} of
{\it conserved} quantities, ``$W$-hair'', that maintain
quantum coherence in the presence of a black hole.
Let us discuss the problem
at the microscopic level, and then
the solution we have found in string theory.
Hawking \cite{hawk} has argued that a path
integral formulation of quantum gravity
would require the use of density matrices
$\rho$ and the absence of an $S$-matrix :
\be
    \rho _{out} = {\nd S} \rho _{in} \qquad : \qquad
    {\nd S} \ne S S^{\dagger}
\label{smatr}
\ee

\noindent This asymptotic loss of quantum coherence would be
related
to a {\it modification} \cite{ehns} of the time-evolution
equation of the density matrix
of a physical system, with hamiltonian $H$,
interacting
with gravity. The modified equation reads \cite{ehns}

\be
   \partial _t  \rho = i[\rho, H] + {\nd  \delta H} \rho
\label{liouv}
\ee

\noindent The equation is similar to that describing time
evolution of open systems interacting with a reservoir
(Markov processes). The
modification (\ref{liouv}) is consistent with the
conservation of probability
\be
 Tr \rho _{in} = Tr \rho _{out}
\label{prob}
\ee

\noindent however, it allows for the evolution of a pure
state into a mixed state
\be
   Tr \rho _{in}^2 \ne Tr \rho _{out}^2
\label{mixed}
\ee

\noindent and it implies that symmetries
do not necessarily lead to conservation laws
\be
   \nonumber \langle A \rangle \equiv Tr (A\rho) \qquad :
\qquad     \langle \partial_t A \rangle \ne 0
\label{symc}
\ee

\noindent In general it leads to the decay of off-diagonal
elements in the density matrix and hence to the collapse
of the wave-function \cite{moh}, with the disappearance
of the interference terms so troublesome to macroscopic
classical intuition.
The classical interpretation of (\ref{liouv}) is the
breakdown of Liouville's theorem which implies the
non-invariance of the phase-space of the matter system
under time evolution.
\pr
In the string theory at hand,
the existence of target space $W$-symmetries seems to
{\it guarantee} the invariance of the matter
phase-space \cite{witt2,emn2}
and hence the extra term  ${\nd  \delta H} \rho$
in the evolution equation (\ref{liouv}) is
{\it zero} due to {\it stringy} symmetry reasons.
\pr
First demonstrated in a two-dimensional model, this
argument applies also to four-dimensional black holes, as
has been shown explicitly in the spherically-symmetric
case of physical interest \cite{emn4}.
To see this, one starts
from the effective action of a generic four-dimensional
superstring theory:
\be
S_{eff}^{(4)}=\int d^4x \sqrt{G} \{ \frac{1}{\kappa^2} R^{(4)}
-\frac{1}{2} (\nabla _{\mu} \phi )^2 -
e^{-2\sqrt{2}\kappa \phi} H_{\mu\nu\rho}^2 +...\}
\label{effect}
\ee

\noindent where $\phi$ is the model-independent four-dimensional
dilaton, $H_{\mu\nu\rho}$ is the antisymmetric tensor
field strength, and the dots represent other degrees of freedom
that are not essential at this stage. We make the
spherically-symmetric
ansatz \cite{emn4}
\be
       ds^2 =
       g_{\alpha \beta} dx^\alpha dx^{\beta} + e^{W(r,t)}d\Omega ^2
\label{ans}
\ee

\noindent where $x^{\alpha}$ are $(t,r)$ coordinates
and $d\Omega ^2=d\theta ^2 + sin^2 \theta d\phi ^2 $, as usual.
An example of (\ref{ans}) is the Schwarzschild metric
in Kruskal-Szekeres coordinates \cite{wheel}:
\be
    ds^2 = \frac{-32 M^3}{r}e^{-\frac{r}{2M}}dudv + r^2d\Omega ^2
\label{krusk}
\ee

\noindent After an appropriate change of variables \cite{emn4}
the radial part of the metric can be mapped onto a two-dimensional
metric of the form
\be
     g^{bh}_{\alpha\beta}dx^{\alpha}dx^{\beta}=\frac{e^D(u',v')}
{1-u'v'} du'dv'
\label{twod}
\ee

\noindent which is nothing other than a conformally-rescaled
form of the known black hole solution in two-dimensional string
theory \cite{witt}.  In general, the spherically-symmetric
ansatz (\ref{ans}) applied to (\ref{effect})
will give a two-dimensional effective action
\be
S_{eff}^{(2)}=4\pi \int d^2x \sqrt{g} e^W (R^{(2)}-(\nabla W)^2
-2 + (\nabla T)^2 + V(T) +... )
\label{efftwo}
\ee

\noindent where $W$ is a new dilaton arising from
the $4 \rightarrow 2$ reduction, the $T$-field is a so-called tachyon,
which is actually massless in two-dimensional string theory, and arises
from
the four-dimensional
matter fields (as a collective representation), and (\ref{efftwo})
contains a two-dimensional cosmological constant that arises from
an integral over the surface of $S^2$. In what follows we shall
study the properties of this effective two-dimensional string theory,
using the particular results derived for $D=2$ stringy black holes
\cite{witt}.
\pr
It has been shown in
general \cite{emn3}
that,
within the context of string theory,
black hole decay can be calculated as a conventional
quantum-mechanical process which does not involve thermal
or other mixed states.
The basic idea is to evaluate the effective action
for strings in black hole backgrounds, summed up
over world-sheet genera. A toy computation
on a torus \cite{emn3}
shows the existence of {\it imaginary}
parts accompanying space-time curvature terms.
In two target-space dimensions
it is easy to miss these imaginary parts
in string theory; the trick is first
to
perform an analytic continuation of
the central charge of the Liouville theory,
so as to compute correlation functions \` a
la Goulian and Li \cite{goul} . This amounts to
dimensionally regularising the target space dimensions $D$
by going to the regime $D$ $>$ $2$. Then the torus
amplitudes have modular infinities
which should be regularised by analytic
continuation \cite{marc}. The imaginary
parts one obtains this way survive the limit
$D \rightarrow 2$ which one takes at the very end.
The generic integral involved in the
computation has the form
\be
   \int _R ^{+\infty} \frac{dx}{x}x^{-\beta} e^{(\alpha + i\epsilon)x}
=[\alpha e^{-i(\pi - \epsilon)}]^{\beta} \int _{-\alpha R - i \epsilon}
^{+\infty} \frac{dx}{x} x^{-\beta} e^{-x}
\label{inte}
\ee

\noindent yielding an imaginary part \cite{marc}
\be
   Im \int _R ^{+\infty} \frac{dx}{x} x^{-\beta} e^{\alpha +i\epsilon}
= \frac{\pi}{\Gamma(1+\beta)} \alpha ^{\beta}
\label{imag}
\ee

\noindent In the two-dimensional string case the imaginary
parts of the torus
correction to the Einstein
term in the effective
action are given \cite{emn3} by the limit $\alpha \rightarrow 0$,
$\beta \rightarrow 0$; the result is $\pi$.
By appealing to the existence of an $S$-matrix
(as well as unitarity arguments) which is
guaranteed by the $W$-hair mentioned above,
one then interprets these imaginary parts
as implying instabilities of the massive
black hole
states, which decay quantum-mechanically
with a rate that is determined by these imaginary
parts, as a consequence of the
usual
`optical' theorem
generalised
to the string case
\cite{turok}.
 The number of pure solitonic states
of a stringy black hole has been estimated \cite{emn4} and shown
to correspond to the number of black hole states found previously
by semi-classical field-theoretic arguments \cite{hawk,bek},
i.e. it is proportional to $M^{2}$. Thus
the entropy
of a four-dimensional black hole is only {\it apparent}, and can in
principle be reduced to zero by measurements of the intrinsically
stringy $W$-hair quantum numbers of these solitonic states.
\pr
The question then arises, how could one imagine
measuring these quantum numbers in practice?
We have
proposed \cite{emnl}
two types of possible measurement.
One is via selection rules for the scattering of light particles
off stringy black holes, and the other is an infinite set of
stringy Aharonov-Bohm effects.
\pr
The concept of the first type of measurement
is very similar to that of $\pi$-nucleon scattering in the
Skyrme model \cite{skyrm}. In that case the $\Delta(1232)$ $3-3$
resonance, for example, is a higher spin (and isospin) soliton
which can be excited if the $\pi$ energy is resonant, and its
decay satisfies certain selection rules. This picture has been
extended
to higher resonances, with soliton calculations
reproducing well the phase shifts in different partial
waves \cite{km} and general selection
rules were
derived \cite{mb}. In our case, there is an infinite
set
of black hole soliton states, classified by the quadratic
Casimir and `magnetic' quantum numbers of an internal symmetry
group, which are excited at calculable energies and decay into
distinctive numbers of light final-state particles. These results
are derived in the limiting case of a flat space-time background
\cite{klepol}
that represents
the end-point of black hole decay, and then
in the generic (potentially macroscopic) black hole
case \cite{emnl}.
The key r\^ ole of the $s$-wave dynamics and the particle
production selection rules are reminiscent of the
Callan-Rubakov \cite{caru} process in scattering off a monopole.
\pr
The second type of measurement \cite{emnl}
involves the characteristic Aharonov-Bohm
interferences between states propagating in the neighbourhood of,
and far from,
a black hole. Since there are infinitely
many massive string states which can be scattered in this way, there
is an infinity of such possible measurements, albeit
with certain practical difficulties in the case of a macroscopic
black hole.
\pr
\section{W-symmetries on the world-sheet
and in physical space-times}
\paragraph{}
We begin our study with
a brief review of
the $W$-symmetries possessed by
string theories
in two-dimensional
space-time, and hence
also
string theories in spherically-symmetric
four-dimensional space-time.
\pr
The first notion of a target space-time $W$-symmetry
was presented by Avan and Jevicki \cite{ava}
in the context of the collective field representation
of the $c=1$ matrix model \cite{das}. The
latter is nothing other than
string field theory for the only propagating degree of freedom
of the two-dimensional strings, the so-called
`tachyon', which is actually massless in such theories.
The existence of an infinite-dimensional Cartan subalgebra
of conserved charges was
demonstrated.
Subsequently,
it was
suggested \cite{emn1} that these symmetries could be understood
as `hidden' gauge
symmetries \cite{ven}
of the underlying
string theory,
associated with the higher excited string states, which in two dimensions
are non-propagating and have definite values of energy and momentum.
These states had been known to exist in intermediate channels
of tachyon scattering amplitudes in matrix models \cite{gross}, in
Das-Jevicki theory \cite{dem},
and in
$c=1$
Liouville
theory
\cite{pol}, but their physical
significance had not been realised at the time.
It is by now clear that these modes
are essential for the perturbative
unitarity of the
flat-space string scattering matrix  \cite{sak}.
In fact, it is through the usual factorisation via the
operator product expansion (OPE)
of vertex operators
that these discrete states are produced as intermediate
states in tachyon
scattering amplitudes.
Soon after the suggestion of ref. \cite{emn1}
on the association of $W$-symmetries with higher-level string states,
Moore and Seiberg \cite{ms}
constructed explicitly a $W_{1+\infty}$ algebra of
symmetries in the fermionic representation of $c=1$ matrix models,
and showed its connection
with the higher-spin (discrete)
string modes. In view of the target-space interpretation
of the model
as a field theory of the tachyon field \cite{das}, these
charges characterise
the flat string background in two dimensions, or the $s$-wave
sector of
dimensionally-reduced four-dimensional strings \cite{emn4}.
\paragraph{}
Although the
target-space interpretation of these symmetries
was evident
by construction,
it is
still
useful
to
understand their origin as
world-sheet symmetries, which would then be elevated to
target-space ones in the usual way, via induced {\it canonical}
deformations
of the corresponding conformal field theory \cite{ovr}.
\paragraph{}
Let us be more precise. String theory in first-quantised form
is formulated as a theory on the world-sheet.
Hence, at a
superficial
level the only apparent symmetries
are those on the world-sheet.
One is
then
interested in knowing under which conditions
such symmetries
could be
elevated into symmetries of the physical
space-time.
Let
$h=\int d\sigma j(\sigma)$ be a conserved charge of
a current $j(\sigma)$ generating such
a symmetry.
This implies
the invariance of the
Fradkin-Tseytlin
generating function for amplitudes under
an appropriate change of the $\sigma$-model fields (target-space
coordinates) that corresponds to the symmetry in question \cite {ven}.
The parameters of the transformation may, and in fact
do in
our case \cite{emn2}, have an explicit dependence on world-sheet
coordinates. This makes this formalism unwieldy for
getting non-perturbative results in some
closed form. As an alternative, Evans and Ovrut \cite{ovr}
suggested the study of hidden symmetries through the induced
deformations on the stress-tensor of the conformal field theory
in a certain background \{ g \}. If $T_g$
(${\overline T_g} $) denotes
the holomorphic (antiholomorphic) part of the
stress
tensor of the $\sigma$-model, then the induced (infinitesimal)
deformation is $ \delta T_g = i[h, T_g]$. The deformation is
a symmetry of the physical (target)
space-time if

\be
    \delta T_g = T_{g+\delta g} - T_g
\label{symm}
\ee

\noindent for some induced transformation of the couplings:
$g$  $ \rightarrow $  $g + \delta g $. A
deformation
is
said to be
{\it canonical} if \cite{ovr}

\be
     \delta T_g = \Phi _{(1,1)}
\label{can}
\ee

\noindent where $\Phi _{(1,1)}$ is a primary field of dimension
$(1,1)$. Thus conformal invariance is automatically satisfied
for canonical deformations. Due to the completeness of the
set of $(1,1)$ vertex operators in string theory, it is also
evident, in view of (\ref{symm}), that a canonical deformation
is also a {\it symmetry} of target space-time.
An important comment is that, if the current generator of the
world-sheet symmetry is an operator of conformal dimension
$(1,0)$ or $(0,1)$, then the induced deformation has conformal
dimension $(1,1)$ and, hence, generates a target space
symmetry of the background. This is precisely what happens in the
case of two-dimensional string theory in flat space-times
\cite{witt2}. There is an infinity of $(1,0)$ or
$(0,1)$
world-sheet currents,
constructed out of states of non-standard ghost number,
which generate $W$-symmetries on the world-sheet
that can be lifted \cite{emn2}
to $W$-symmetries of the
physical space-time
of two-dimensional string theory, formulated on
flat
backgrounds. These symmetries have the property
that
they leave invariant
under
target
time evolution
the two-dimensional
tachyon
phase-space,
which, in view of \cite{das},
is
equivalent to a matrix-model phase space \cite{witt2}.
Following
indications about
background independence of the Das-Jevicki string feld theory
\cite{das,alw,emn2},
one might expect \cite{emn2}
that
these $W$-symmetries, or some appropriate
deformations of them,
would persist in highly-curved string backgrounds such as
black holes \cite{wad,witt}.The phase-space-area-preserving
character of the symmetries, then, would guarantee the
maintenance of quantum coherence even during extreme
 physical
processes
like black hole evaporation/decay, for the
reasons
argued in \cite{emn2}.
\pr
To check these considerations explicitly for curved
space-times seems a difficult task. The
weak-field
expansion methods that are used in
extracting the $\beta$-functions of
$\sigma$-model
perturbation theory might not prove sufficient
for getting exact information about
complicated backgrounds
such as black holes.
On the other hand, already
from critical string theory
we know
of cases where exact
conformal field theories
have been used to
circumvent the patterns
of perturbation
theory
around complicated string ground states,
like Calabi-Yau compactified
spaces. Gepner's construction \cite{gep}
of tensoring superconformal field theories
has provided us with non-perturbative information
about {\it exact} solutions of string theory
in the form of Calabi-Yau Ricci-flat spaces,
which
seemed not to correspond
to solutions of perturbative $\beta$-function
equations \cite{pope}.
In Gepner's construction an important method
was the use of Fateev-Zamolodchikov
parafermions
to represent
the pertinent
$N=2$ superconformal algebras. It is in this
representation that {\it extra} selection
rules, not known previously,
have been found for correlation functions in
Calabi-Yau backgrounds \cite{ross}.
\pr
Motivated by these results in critical string theory,
it was
natural to ask
whether
similar constructions of exact conformal field theories
can be found in the case of
string
black hole backgrounds. This question was answered positively
by Witten \cite{witt}, who
has shown that it is possible to describe the
interior of the horizon of a black hole in two-dimensional
space-time using a coset $\frac{SL_k (2,R)}{G}$ Wess-Zumino(WZ)
model, with $G=SO(1,1)$ ($G=U(1)_{compact}$) for Minkowskian
(Euclidean) black holes. The important point is that
this is the first time that a {\it finite} theory, like the WZ model,
is used
to describe a target space-time singularity
\footnote{However, it should be stressed
that
the conformal field theories
in question are not a mere ansatz for describing
known local field theory objects
like traditional
Schwarzschild solutions of Einstein's equations.
On the contrary, they serve
to
demonstrate
the fact
that objects
{\it resembling} the space-time singularity structure
of spherically-symmetric non-rotating
four-dimensional black holes \cite{emn4}
appear
as {\it exact} solutions of subcritical
string theory \cite{aben}. This is
similar in spirit to
Gepner's demonstration
that Calabi-Yau (Ricci-flat) spaces are exact (non-perturbative)
solutions of critical string theory.}.
And it is this kind of singularity
that thwarted
earlier attempts to
describe
black holes, or in general
singular objects surrounded by event horizons, in a way
consistent with quantum mechanics \cite{hawk}. Conformal
field theory provides, as we shall
see in section 3, the
consistent construction of a scattering
matrix to describe scattering of
propagating string states (tachyons)
off the black hole background. This is to
be contrasted with the
situation
in point-like field theories,
where such a scattering
matrix was argued not to exist \cite{hawk,ehns}.
\pr
In parallel to
the parafermion representation of Gepner's spaces,
one
could
expect a similar construction
here, as was indeed
shown
in \cite{dixon}. Any state that admits an expansion in terms
of an $SL(2,R)$ current-algebra basis
can be represented as a highest weight
$N=2$ superconformal state,
which admits a parafermion realisation \cite{lyk}.
In bosonised parafermion language, an $SL(2,R)$
algebra is realised by three bosons. To
describe a coset requires factoring out one of them,
and one is left with a two-boson realisation
of the coset model. In complex notation, let $\phi (z)$,
${\overline \phi (z)}$ be the two bosons, with
$j = \partial _z \phi$, $ {\overline j}= {\overline \partial _z \phi}$
the corresponding currents. The parafermion currents are
expressed as
\be
\psi _{+}={\overline j} e^{\phi + {\overline \phi}},
\psi _{-} = j e^{-(\phi + {\overline \phi} )}
\label{paraf}
\ee

\noindent The $W$-symmetry structure is revealed
by
looking at the classical
OPE
between parafermion currents $\psi _{+}(z)\psi _{-}(w)$,
for $z \rightarrow w$ \cite{kir}.
Expanding in powers of $\epsilon \equiv z-w$
and keeping {\it all} powers of $\epsilon$
\be
   \psi _{+}(z)\psi _{-}(z + \epsilon) =
\sum_{r=0}^{\infty} u_r (z) \frac{\epsilon ^r}{r\!}
\label{wsymm}
\ee

\noindent one discovers a world-sheet $W$-algebra,
generated by the commutation relations of the
$u_r$ \cite{wu,wu2}.
Parenthetically, but importantly,
we notice that the coefficients $u_r$
are world-sheet currents of conformal spin $r+1$ \cite{wu2},
which
appear as expansion coefficients
in the
pseudo-differential
operator
$L=\partial _z +\sum_{r=0}^{\infty} u_r (\partial _ z)^{-r-1}$,
defining
the so-called
$KP$
Hamiltonian
basis. In this sense the appearance
of $W$-symmetries is related to the $KP$ hierarchy,
which might be suggestive of new ways of
approaching $c=1$ string theory \cite{wu2}.
In the coset model, the
symmetry algebra generated by the currents $u_r$
is related to the second Hamiltonian
structure of the
$KP$ hierarchy \cite{wu}.
It is actually
a non-linear deformation \cite{wu2}, ${\hat W _{\infty}}$,
of the centerless $w_{\infty}$ algebra
of Bakas\cite{bak}. We shall come
back to this point in section 4. Quantisation
of the model does not simply
require
{\it normal ordering}
these
OPEs, but a {\it redefinition} of
the currents
\cite{wu} so as to ensure the closure of the
quantum ${\hat W}$-algebra. Details of the
construction can be found in \cite{wu}.
\pr
An infinite set of commuting quantum $W$-charges is
constructed as world-sheet spatial integrals
of the
currents $u_r$ (or rather an appropriate redefinition, $W_r$,
in the notation of \cite{wu2,kir}),
with
integer
conformal spin
$s \ge 2$ \cite{wu}
\be
   [ \int dz W_s(z), \int dw W_{s'}(w)]=0 ; s,s' \ge 2.
\label{charges}
\ee

\noindent The charge $W_2$ coincides with the
Hamiltonian of the model, or,
in the case of closed strings,
with the holomorphic part of the $L_0$ Virasoro operator.
Notice also
that spin-one objects
are not included, by construction \cite{kir,wu},
in this set and so the
deformed algebra is not of $W_{1+\infty}$ type
(which is generated by integer
conformal spins $s \ge 1$). The $W_{\infty}$
algebra is known \cite{wu2} to be a subalgebra
of $W_{1+\infty}$. However, it is the $w_{\infty}$
algebra
that has
a well-known geometrical
interpretation as a
phase-space area preserving
symmetry \cite{bak}, and so for our purposes
\footnote{It should be
mentioned,
though, that in view of the matrix-model
result \cite{ms}, one might expect
the actual
target space-time symmetry of the two-dimensional
string theory to
be bigger, allowing for
symmetries generated by conformal spin-one currents.
This would include the canonical deformations
mentioned above. Of course, one cannot exclude the possibility
that the flat space-time symmetries are
larger
than those of the black hole background.}
it is
sufficient to concentrate on $w_{\infty}$ and its
quantum deformation ${\hat W} _{\infty} $.
\pr
The above constructions are valid for coset WZ models
whose level parameter $k \ge 2$ \cite{kir,wu}.
The model with $k=\frac{9}{4}$ admits an interpretation
as a {\it critical} string theory propagating in a black
hole background \cite{witt}.
The interesting point is therefore the
lifting of this enormous
world-sheet symmetry to a
physical gauge symmetry of two-dimensional
strings. Due to the higher conformal spins of the
generating
currents, it turns out that the induced deformations
about the pertinent background are not in general
canonical. As argued in \cite{gian} recently,
canonical deformations are only a part of the
enormous infinite set of target gauge symmetries.
The requirement that
a field
be $(1,1)$
implies usually an equation of motion {\it and }
some gauge conditions (constraints), for
higher-spin states.
The most general set of deformations discussed
in \cite{gian} relaxes the requirement of
gauge fixing, and hence one
works in arbitrary gauges, so the only remaining
constraint is
conformal invariance. The main conclusion then is
that
any symmetry on the world-sheet, generated
by a current of arbitrary conformal spin, can be
viewed as generating a symmetry of target space-time,
provided translational invariance is maintained on the
world-sheet. Therefore
the corresponding charge
operator should commute with $L_0 - {\overline L_0}$:
\be
   \{ h \} : [h, L_0 -{\overline L_0}] =0
\label{set}
\ee

\noindent The cost
one pays is the introduction of auxiliary
fields
that are pure gauge artifacts
introduced to count
correctly
degrees of freedom \cite{gian}. In the two-dimensional
case,
the
existence of discrete modes in higher spin levels,
which are associated with
the target $W$-symmetries, emerges precisely from
a relaxation of constraints and/or gauge conditions
which occurs at particular values of energy and momentum
\cite{klepol,pol}. Moreover,
the world-sheet charges (\ref{charges}) do satisfy
(\ref{set}) by construction, and therefore they are
responsible for the generation of string gauge
symmetries in physical space-times. In the context
of first-quantised string theory, these symmetries
will be expressed through complicated redefinitions
of the $\sigma$-model fields \cite{emn2}.
{}From the commutation relations (\ref{charges})
it becomes clear that the induced deformations vanish
when integrated
over the
world-sheet. This
should be intuitively expected. The ordinary
target-space general coordinate transformations
are in fact {\it total} derivative effects on the
world-sheet,
being mainly responsible for the difference of
local from global conformal invariance (c.f. the
difference
of $\beta$-functions from ${\overline \beta}$-functions
is expressed as
a general coordinate transformation \cite{tsey}).
Also,
from
a glance at the commutation relations of the charges
(\ref{charges}) \footnote{The charges are expressible
in terms of $SL(2,R)$
currents and their derivatives,
which have a definite action
(as ladder operators) on a state with quantum numbers
$(j,m)$ \cite{wu}.} with the string level operator in the
corresponding string theory, it becomes evident
that for many of them there is a non-vanishing
(and rather complicated)
result,
thereby implying a mixing of string levels, as expected
for a stringy gauge symmetry transformation.
\section{Measuring W-hair}
\paragraph{}
We have proposed \cite{emnl}
two classes of gedanken experiments
for measuring the $W$--hair.
One involves scattering a light particle
off a black hole. As its energy is varied,
like the pion in scattering experiments
off protons, resonant (Breit-Wigner type)
states with different values of $(j,m)$
are excited with {\it calculable} $S$-matrix
elements. There are selection rules for the
number $N$
of light particles produced when each black hole
state decays \cite{emnl}. These selection rules
are
analogous to those of pion-proton scattering \cite{km,mb}
in the framework of a
Skyrme model for the nucleon.
Details on the derivation
can be found in \cite{emnl}.
Here we only state the result:
\bea
\nonumber     j=\frac{1}{3}m - 1 + \frac{1}{2} N     \\
        j \ge \frac{1}{4} (N-3)
\label{selbh}
\eea

\noindent Notice the correspondence with the
selection rules for the scattering of `tachyons'
in the case of $c=1$ flat space-time strings
\cite{klepol} upon making
the analogy
\bea
\nonumber     2j_{SL(2,R)} \rightarrow j_{SU(2)}-1  \\
              \frac{2}{3}m_{SL(2,R)} \rightarrow m_{SU(2)}
\label{sutwo}
\eea

\noindent However,
(\ref{sutwo}) should only be considered
as a formal correspondence. The set of discrete states
in coset theory is larger than that of flat $c=1$ strings
\cite{distl},
and hence there are always {\it extra} selection rules in
the former model.
\pr
The above selection rules have been originally
derived for the case of the Euclidean black hole .
To pass into Minkowski formalism one notices the following.
{}From the cohomological approach of Distler and Nelson \cite{distl}
it becomes clear that the higher-level discrete states
in this string theory are characterised by {\it real j}.
The usual passage from Euclidean to Lorentzian black holes
is then given by the replacement $m \rightarrow i\mu$,
where $\mu$ is defined by the diagonalisation of the generator
of the $SO(1,1)$ subalgebra in the coset model
$\frac{SL(2,R)}{SO(1,1)}$.
For the discrete states $\mu$ is taken to be purely imaginary,
$\mu=-im$, so both  selection rules (\ref{selbh}) can be
translated directly into the Minkowskian theory.
Again, however, in view of the {\it inequivalent} spectra
of the two theories the physical
content of the (formally identical)
selection rules changes.
\pr
The physical interpretation of these selection rules is
clear.
They
constitute a manifestation of the
underlying $W$-symmetry structure of the theory.
The
non-trivial conserved charges
that characterise the background (black hole)
play a
r\^ ole
analogous
to {\it topological }
charges in conventional local field theories, labelling
not particles but rather vacuum sectors of the theory
where
the scattering of particles takes place. The key r\^ ole
of the
$s$-wave dynamics and the selection rules are
reminiscent of the Callan-Rubakov \cite{caru}
process in scattering off a monopole. Further, we
recall that
the instanton
number
in the
standard electroweak model
also leads to selection rules
restricting the number as well as the
flavour
of emitted particles, for
given
incoming states \cite{ringw}. In our case, since
there is only
one flavour of particle,
the only
restriction concerns the number with which the exchanged state
can interact. In contrast to the local field theory
example, however, here we have an {\it infinite} set
of selection rules that differ for each internal state, and an
{\it infinity}
of topological (conserved) charges which label
the internal degrees of freedom of the black hole.
It is this feature
that explains naturally the large statistical
entropy of the latter, with the entropy
being defined \cite{emn4}
as the loss of information for an observer at spatial
infinity  who, ignoring the higher excited black hole states,
measures only the classical energy and charge of a non-rotating
black hole. However, these can in principle be distinguished,
for example, by measuring the number of particles the black hole
states
emit when they decay.
\pr
We discuss now
another class of distinguishing
measurements which involves an infinite set
of Aharonov-Bohm phase effects.
It has been argued \cite{klepol,emnl} that the discrete
states appearing both in the flat space $c=1$ string
and in the black hole coset model are {\it singular}
gauge transformations (or physically equivalent
representations thereof). This picture is consistent
with that stemming from ordinary strings,
where the various states are viewed as gauge particles
\cite{grosmend}. In higher-dimensional string theories,
the fact that
stringy symmetries {\it mix} the various mass levels implies
{\it spontaneous} breaking of these symmetries.
In two dimensions spontaneous breaking cannot occur
\cite{col},
so the gauge $W$-symmetries
constitute a set of {\it unbroken} string gauge symmetries.
The selection rules discussed above are a
manifestation
of this. In topologically non-trivial backgrounds, like the
Schwarzschild black hole, these gauge symmetries
lead to an infinity of
{\it non-trivial} conserved charges. The association
of the latter with discrete (non-propagating) delocalised
string states implies their {\it topological} nature.
Formally this can be seen as follows. At
the level of the target-space effective action
the coupling of the (gauge) string states with
the conserved currents $J^{M...}$
that characterise the string background
can be represented generically as
\be
      \int _{space-time} \sum_{s=string-states}
      A_{M...}^{s} J^{M...}_{s}
\label{coupl}
\ee

\noindent In the
spherically-symmetric case the integral over
the four-dimensional space-time reduces to a two-dimensional
integral. In that case,
the string states $A$ are singular gauge
transforms $d\Lambda$,
and the currents $J$ are induced by
the world-sheet currents leading to the charges (\ref{charges}).
{}From
current
conservation it follows immediately that
the terms (\ref{coupl}) are purely {\it surface} terms,
and therefore they can only carry {\it global} (topological)
information about the space-time manifold.  This is consistent
with the fact that these charges are {\it exactly} conserved
in highly-curved space-times. In fact it is the only possibility.
\pr
The pure gauge nature of the string states that couple to the
conserved quantum numbers of the black hole leads to a better
analogy with ordinary Aharonov-Bohm  experiments used
to measure
conventional discrete gauge hair in local field theories \cite{lah}.
Take, for instance, the antisymmetric tensor hair in string-inspired
black-hole theories
involving gauge fields that couple to the
antisymmetric tensor field strength. There
{\it both} the gauge field strength
and the antisymmetric tensor field strength are
identically
{\it zero} outside the horizon. Also the charge
corresponding to the antisymmetric
tensor field strength is a
surface term \cite{lah} $\int _{S(V)}B$, over
a two-dimensional
surface $S(V)$ surrounding a three-dimensional spatial
volume $V$.
A similar thing happens here. The field strengths
corresponding to the string states vanish outside the
horizon, since the states are pure gauge transforms,
with singularities at the origin of the black hole.
The corresponding charges, characterising the black hole solutions,
are also surface terms. Due to spherical symmetry, the
charges are just conserved numbers, since surface terms
in effectively two-dimensional spaces consist just of points.
The surface character of the charges can be easily exhibited
for the first few of them (e.g. the total energy of the black hole
\cite{witt,emn1}).
That these charges characterise
the spatially-asymptotic form of the coset model is another
manifestation of this fact.
\pr
The Aharonov-Bohm
process for measuring the $W$-hair
will be
similar to that of
antisymmetric tensor hair \cite{lah},
but
with some crucial
differences. Consider the scattering of four-dimensional
fundamental strings off a spherically-symmetric
black hole background.
One considers an interference experiment at a final
point $B$
between fundamental strings that have
travelled from an initial point $A$ via two paths: in
flat space,
and
in a black
hole environment.
The world-sheet of the latter
fundamental string
encloses
the black hole singularity at a
point $C$, and
this produces a surface-like
coupling with
the
correponding charges of the black hole of the form (\ref{coupl}).
The latter being topological
will only manifest itself as an Aharonov-Bohm phase in the
wave function of the fundamental string , which can be measured
via an
appropriate interference experiment at the point $B$.
Due to the fact that the charges in the black hole background are
in correspondence \cite{ms} with the higher-spin states, in
order to measure individual charges one has to use
{\it polarised} fundamental strings where the particular
string mode is excited. Here comes an important
difference from ordinary strings. If we were living strictly
in two dimensions,
these modes could not propagate,
due to their
topological nature, so scattering
these modes
off the
black hole
is not possible in the ordinary sense.
However,
in four-dimensional topologically non-trivial
spherically-symmetric space-times
(whose singularity structure is described by two-dimensional
strings \cite{emn4})
one can consider propagating higher-level string states,
whose $s$-wave sector resembles that of two-dimensional strings,
symmetrywise. In this sense one considers a fundamental
string whose world-sheet
encloses
a black hole at $C$. The fundamental string
is polarised so that
a
level $N$ state is excited, say. The $s$-wave sector of these
states (which is topological) couples to the conserved charges of the
spherically-symmetric black hole and makes a non-trivial phase
contribution to the wave function of the propagating
four-dimensional string at $B$.
\pr
A final comment concerns the formal character
of the process. For the
string propagating in a flat
space-time background
one can
use the $c=1$ string theory to describe its
propagation. However, in this
case there will be a
mismatch when one couples the two theories at $C$,
due to the extra states
that the coset model (black hole background) appears to have
\cite{distl}. Hence the most correct treatment, consistent
from a string field theory point of view, would be to treat
the flat space-time as an {\it asymptotic region} of the
black hole space-time. However,
in view of the similarity
of the target space symmetries between the two models, it
is
possible that the non-trivial
conserved charges are in {\it correspondence} with the
set of states of the asymptotic
$c=1$ Liouville matter system. The existence of stringy symmetries
in the coset model {\it mixing} levels
and {\it mapping}, for
instance, the
discrete states ${\tilde D}^{\pm} \rightarrow D^{\mp} $
\cite{distl} offers
support to these arguments, and
probably implies
that the
actual (physical) spectrum of the coset model might be smaller
than it appears.

\section{Space-time singularities, topological field theories
and $W$-hair }
\pr
We have seen in the previous sections how quantum coherence is
maintained by
spherically-symmetric stringy black holes, by reducing the problem to
one which
is effectively two-dimensional, and then using the known W-symmetry of
the
two-dimensional system. We have also given in \cite{emn4}
reasons why this
analysis
should also be applicable to more general axi-symmetric black holes in
four
dimensions. However, it would clearly be nice to find a broader
framework
capable of treating more general singularities, including the initial
singularity of the standard Big Bang cosmology.
\pr
An interesting possible approach to this problem has recently been
proposed
by Eguchi \cite{eguchi}. He has
shown that the conformal field theory action close to
the
singularity can be rewritten in the form:
\be
S=-\frac{k}{4\pi} \int d^2x D_i a D_i b + i \frac{k}{2\pi}
\int d^2x w \epsilon^{ij}F_{ij}
\label{topol}
\ee

\noindent where
the singularity is parametrized by the limit $uv = 1$
of the
Kruskal-Szekeres coordinates $u$ and $v$ \cite{witt}
which have been written in
the forms
$u = exp(w), v = exp(-w),$ and $a, b$
are diagonal elements in the general
$2 \times  2$ $SL(2,R)$
matrix - the off-diagonal elements being $u$
and $-v$
with $ab + uv =1$. In
equation (\ref{topol}), $F_ij$
is the field strength of the $U(1)$ gauge potential
$A_i$, and
the indices $i,j$ take two values, corresponding to the two dimensions
of the
space-time, and the two variables $a$
and $b$ both vanish at the
singularity. The
exciting and key observation of Eguchi was that the residual theory
resembles a
{\it topological field theory}, namely
the dimensionally-reduced form of a
Chern-Simons theory in three dimensions.
\pr
That a topological field theory should emerge at the singularity was
perhaps
not surprising for at least one reason. It is thought that topological
field
theories correspond to an unbroken phase of gravity, in which the metric
vanishes, and it is precisely at the singularity that there is no
physically-meaningful metric. Here there is an analogy with the core of
magnetic monopole, where the vacuum expectation value of the Higgs field
vanishes at the centre of a topologically-non-trivial solution of the
field
equations, and the underlying gauge symmetry is restored.  In the same
way, we
expect that a non-zero value of the metric is at least one of a possibly
infinite number of order parameters that mark the spontaneous breakdown
of a
higher stringy symmetry, which should become manifest in the associated
topological field theory.
\pr
As described in previous sections, we have identified the essential
symmetry
that safeguards quantum coherence as a $W$-symmetry, with
an infinite set
of
associated conserved charges that are in principle measurable. Indeed,
there is
evidence that the topological field theories relevant to space-time
singularities
have a high degree of symmetry that includes this
$W$-symmetry.
Eguchi and Yang
\cite{egyan}, based on earlier work by Witten \cite{witt3},
have given a prescription for constructing such theories by
twisting $N = 2$
superconformal Wess-Zumino models. Such models have recently been
considered in
the literature in connection with a supersymmetric version of the
two-dimensional black hole \cite{noij}. The
model is described by an action of
the form:
\be
    S_{susy}= S_{WZ} +
    \frac{i}{2\pi} \int d^2z (Tr \psi D_{{\bar z}} \psi
+ Tr {\bar \psi} D_z {\bar \psi})
\label{susy}
\ee

\noindent where $\psi$, ${\bar \psi}$ are fermions of the coset
$\frac{SL(2,R)}{U(1)}$ in the case of Euclidean black holes.
The $N = 2$ supersymmetry is twisted by adding a
multiple of
$\partial _z J$, where $J$
is the $U(1)$ current, to the stress tensor of the
conformal field
theory. The resulting twisted thory has vanishing central cherge with
respect to
the redefined stress tensor, which can be thought of as a $BRST$
commutator. The
fermions of the superconformal model (\ref{susy})
become ghost fields of the
topological
field theory. Eguchi has shown that the path-integral of this $N =2$
superconformal Wess-Zumino model, interpreted as a topological field
theory,
describes observables in the neighbourhood of the singularity.
It
remains an
interesting problem to establish a dictionary relating the calculable
quantities
of this
theory,
namely
the Wilson loops, to observables in black hole
physics.
Since they are non-local quantities, it is tempting to speculate that
they are
related to the non-local $W$-charges.
\pr
Upon
twisting the $N=2$ supersymmetry, the singularity
becomes the fixed point of the $BRST$-transformation
which is the usual condition for defining
{\it physical} states
\be
  Q|physical\rangle =0
\label{brst}
\ee

\noindent where $Q$ is the $BRST$ operator. In bosonic
two-dimensional string theory the $BRST$ cohomology
of physical states includes states with non-standard
ghost number, which are actually responsible for the
$W$-symmetries of the model \cite{witt2}.
\pr
In the case of topological field theories one would expect
a similar situation.
Before twisting, the bosonic sector of the model
(\ref{susy}) is an ordinary
Wess-Zumino theory that
describes a black hole space-time
with an infinite set of quantum hair, due to the
${\hat W}_{\infty}$ symmetry
discussed
in section 2 \cite{emn1,emn2,kir,wu}.
The
generators of the symmetry appear in the higher
orders of the
$OPE$ between
parafermions of the form (\ref{paraf}).
The coefficients of that expansion are the same
as those of the $KP$ pseudo-differential operator
\cite{wu2}
\be
     L=\partial _z + \sum_{r=0}^{\infty} u_r (\partial _z)^{-r-1}
\label{kpd}
\ee

\noindent The latter can indeed be expressed in terms
of bosonic currents $j(z)$ and their derivatives \cite{wu},
and it is in this way that non-linear deformations in
the $W_{\infty}$ algebra arise in these models.
\pr
The exciting observation from our point of view is that the $N = 2$
superconformal Wess-Zumino models are known to possess a supersymmetric
extension of $W$-algebra, which actually contains two independent
$W_{\infty}$
algebras, associated with the
components of the
superfield version of the
$KP$-hierarchy:
\be
    \Lambda = \theta ^2 + \sum_{r=0}^{\infty}U_{r} \theta ^{-r-1}
\label{skp}
\ee

\noindent where $\theta \equiv \partial _{\zeta} + \zeta \partial _z$
is the supercovariant derivative
in (1+1)-dimensional superspace, and the superfields
$U$
have the following component expansion
\be
     U_r=v_r(z) + \zeta u_r(z)
\label{ucoe}
\ee

\noindent The two sets of coefficients $u$ and $v$ generate
(after appropriate redefinitions \cite{yu}) two
supersymmetric
$W$-algebras
without centre. The fact that the two sets of coefficients
(\ref{ucoe}) {\it commute} with each other
implies that the bosonic sector of the algebra generated by
(\ref{skp}) has the form of a direct sum of $W$-algebras
\be
 W_{1+\infty} {\bf \oplus } W_{1+\infty}
\label{sum}
\ee

\noindent which is called
{\it super}-$W_{1+\infty}$.
The existence of a direct
sum of symmetry
algebras
in the bosonic sector is a general feature
of the supersymmetric extension of $W$-symmetries \cite{sezg}.
\pr
{}From the analogy
with the bosonic case, one expects that by looking at the
second Hamiltonian structure of the super-$KP$ operator
(\ref{skp}) one should obtain a direct sum
$W_{\infty} {\bf \oplus } W_{\infty}$ algebra (in the bosonic sector)
which is contained \cite{wu2} in the super-$W_{1+\infty}$.
This would
imply
that the {\it symmetry group} that
characterises $N=2$
supersymmetric models of the type (\ref{susy})
would be at least
${\hat W}_{\infty} {\bf \otimes} {\hat W}_{\infty}$.
This would
confirm
our expectation that the topological
field
theory should have a symmetry higher than the $W$-symmetry of the
two-dimensional
black hole solution. It
is tempting to speculate that there may be an
analogy
between the breaking of $W$ $\otimes$ $W$ $\rightarrow$ $W$
via a non-singular
value of the metric, and
possibly other order parameters, and
the breaking of chiral $SU(N)  \otimes  SU(N)$
$\rightarrow$ $SU(N)$ symmetry
via a scalar condensate.
\pr
The continuation of this line of research requires a generalization
of the
construction of Eguchi to other $N = 2$
superconformal Wess-Zumino models
and
topological field theories able to describe other space-time
singularities of
physical interest. Until this is done, the relevance of this approach to
such
important extensions of our black hole studies as the elucidation of
space-time
foam should be considered as a speculation. However, it
may well be that some Wick rotation of
Eguchi's
construction could describe the initial singularity of Big Bang
cosmology. To
the extent that there is no metric in the topological phase close to the
singularity, and quantum coherence is maintained by some $W$-symmetry,
permitting
apparently superluminal Einstein-Podolski-Rosen correlations, these
ideas might
provide a resolution of the cosmological horizon problem.

\newpage
\noindent {\Large{\bf Acknowledgements}} \\
\par
The work of D.V.N. is partially supported by DOE grant
DE-FG05-91-ER-40633.
\pr

\end{document}